\documentclass[final,5p,times]{elsarticle}
\usepackage{bbding}
\usepackage{pifont}
\usepackage{soul}
\usepackage{amssymb}
\usepackage{mathrsfs}
\usepackage{amsmath}
\usepackage{enumitem}
\usepackage[ 
colorlinks=true,
pdfborder={0 0 0},
linkcolor=red,
citecolor = red,
breaklinks
]{hyperref}
\usepackage{xurl}
\usepackage{graphicx}
\usepackage{placeins}
\usepackage{physics}
\usepackage{bm,accents}

\usepackage{dcolumn}
\usepackage{xcolor}
\soulregister\citep7
\soulregister\citet7
\soulregister\pageref7

\usepackage{booktabs}
\biboptions{numbers,sort&compress}
\journal{Physica A: Statistical Mechanics and its Applications}
\makeatletter

\newcommand{\Rmnum}[1]{\expandafter\@slowromancap\romannumeral#1@}
\makeatother
\usepackage{caption}

\begin{document}
\begin{frontmatter}

\title{Phase diagram and transition properties of the Blume-Capel model: A Monte Carlo study}

\author[aff1]{Tenglong Li}
\author[aff2]{Zhigang Feng}
\author[aff1]{Jie Zhang\corref{cor1}}
\cortext[cor1]{Corresponding author}
\ead{zhangjie01@tyut.edu.cn}

\address[aff1]{College of Physics and Optoelectronics, Taiyuan University of Technology, Shanxi 030024, China}
\address[aff2]{National Institute of Metrology, Advanced Measurement Science, Beijing 100029, China}

\begin{abstract}
 The $J_x$-$J_y$ Blume-Capel model, with and without an external magnetic field, is investigated using the Monte Carlo method. Our aim is to verify the location of the tricritical point, examine the existence of reentrant behavior, and study the magnetocaloric properties of this
 specifically designed system. In the non-field case, the critical temperature given by the Monte Carlo simulations is lower compared to the conventional antiferromagnetic square lattice. When an external magnetic field is applied, we observed signs of a one-dimensional spontaneous symmetry‑breaking phase in the mixed-states regime, which is counterintuitive with secondary specific-heat peak. We take the snapshot of this state and elucidated the underlying mechanisms using mean field theory in the field of spin-1 spinor Bose-Einstein condensate . Regarding reentrant behavior, we confirmed its occurrence as the magnetic field varies, but only when the temperature is kept constant. Temperature-dependent reentrant behavior could not be identified within this system. Finally, we examined the magnetocaloric properties and found no evidence of first-order phase transitions within the region of the considered phase-transition point. Although both inverse and direct magnetocaloric effects are present and related to the mixed state, we identified that the relevant phase transitions are all of the second order.
\end{abstract}

\end{frontmatter}

\section{Introduction}
The Blume-Capel (BC) model, introduced independently by Blume~\cite{Blume} and Capel~\cite{Capel} in 1966, has provided a widely used statistical framework for studying spin-1 critical phenomena and phase transitions in magnetic systems. The model incorporates a vacant zero-spin degree of freedom that is absent in the conventional binary Ising model, and includes a zero-field splitting (crystal field) term in the Hamiltonian, which favors the zero-spin vacancy state~\cite{vacancy}.
The key feature of this system is the tricritical point (TCP)~\cite{Mataragkas2025Transfer}, which separates the first-order and continuous segments of the ordered-to-paramagnetic phase boundary.

Over the years, the BC model and its closely related spin systems remain active research topics and play important roles across condensed matter physics, statistical mechanics, and materials physics~\cite{Akkurt2026Geometry,Vatansever2026Skyrmion,Ye2026Ferromagnetic,fer3/2,Shu2026Metallic,Liu2026Magnetocalorics}.
The development of this field is multidimensional, involving anisotropy~\cite{Fytas2018Disorder,Gulpinar2011Diluted}, spin-glass generalizations~\cite{Lyakhov2025SpinGlass}, antiferromagnetic spin-$3/2$ system~\cite{anti3/2}, and machine-learning methods~\cite{Babu2025MachineLearning}. Other extensions include random crystal fields~\cite{random1spin} and competing short- and long-range interactions~\cite{shortandlong}. Beyond modifications of the Hamiltonian, research has also addressed different geometries, including triangular-lattice BC models~\cite{triangular,Mataragkas2025Transfer} and the three-dimensional BC model~\cite{three-dimensional}. Despite its long-standing theoretical importance, exact solutions are available only in restricted cases, and estimates of critical and tricritical behavior can depend on dimensionality, lattice geometry, and the approximation employed~\cite{Blume,Capel,2018blume,meanfieldtheory2022,variationalapproach,effective-field}. This implies that the influence of different geometric structures on the tricritical point remains incompletely understood. 

\begin{figure}[!t]
\centering 
\includegraphics[width=0.65\linewidth]{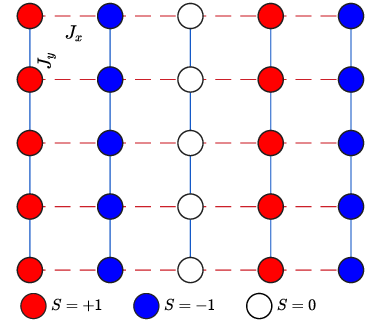}
\caption{Square-lattice $J_x$-$J_y$ Blume-Capel model. Red, blue, and open circles denote $S=+1$, $-1$, and $0$, respectively. The horizontal dashed bonds represent the $J_x$ interaction, while the vertical solid bonds represent the $J_y$ interaction.}
\label{lattice_model}
\end{figure}

In this paper, we focus on a specifically designed $J_x$-$J_y$ BC model on a square lattice. The exchange couplings $J_x$ and $J_y$ act along the $x$ and $y$ directions, respectively; for the main parameter set considered below, $J_x<0$ is antiferromagnetic and $J_y>0$ is ferromagnetic~\cite{PH.A1,JXJY-Ising}.
This work is motivated by two key considerations: (1)
Ruby artificial spin ice provides an experimental realization of BC three-state degrees of freedom through toroidal moments, thereby connecting statistical lattice models with an engineered condensed-matter platform~\cite{Berchialla2026RubyASI}. The demand for specially designed systems and corresponding theoretical predictions has grown. (2) The Cluster Variation (CV) Method has been applied to this $J_x$-$J_y$ BC model~\cite{PH.A1}, predicting a superantiferromagnetic (SAF) phase and a mixed phase between the SAF and paramagnetic phases. Predictions such as first-order transitions, reentrant behavior, and magnetocaloric response~\cite{PH.A1,PH.A2},  urgently require verification using independent approaches.
Considering the influence of an external field, the total Hamiltonian of the $J_x$-$J_y$ BC model can be expressed as follows:
\begin{equation}
H=-J_x\sum_{\langle i,j\rangle_x}S_iS_j-J_y\sum_{\langle i,j\rangle_y}S_iS_j+D\sum_i S_i^2-h\sum_i S_i,
\label{eq:ham1}
\end{equation}
where $D$ is the anisotropy parameter or crystal-field term, and $h$ represents the external magnetic field. The spin variable $S_i$ takes the values $-1$, $0$, and $1$ at site $i$. The sums over $\langle i,j\rangle_x$ and $\langle i,j\rangle_y$ run over nearest-neighbor pairs along the $x$ and $y$ directions, respectively. The sample configuration and basic constructions can be found in Fig.~\ref{lattice_model}.

To better understand the phase-transition properties for $J_x=-J_y$ interactions, we carried out Monte Carlo (MC) simulations to reconsider several major conclusions for this model. The advantages of the MC method are obvious: (\uppercase\expandafter{\romannumeral1}) In the CV method, the cluster size is often limited; however, this constraint does not apply to the MC method. (\uppercase\expandafter{\romannumeral2}) While analytical methods may struggle to effectively address multi-spin systems, the MC method can capture these effects more comprehensively. (\uppercase\expandafter{\romannumeral3}) The MC method allows for the direct visualization of the system's configuration, enhancing interpretability.
(\uppercase\expandafter{\romannumeral4}) Under adiabatic conditions, changing the applied magnetic field can change the temperature of a magnetic material.Because the Metropolis algorithm samples the Boltzmann distribution, MC simulation is naturally well-suited for studying temperature-dependent effects.
The outline of this paper is as follows: In Section~\ref{sec:model}, we introduce the observed quantities related to our investigation. Section~\ref{sec:res} presents our main results, wherein we compare our findings with the analytical conclusions derived in~\cite{PH.A1,PH.A2} and discuss an interesting finding in the mixed phase. Additionally, we offer insights into significant phenomena, like the reentrant behavior and the magnetocaloric effect. Finally, we conclude with a summary of our findings in Section~\ref{sec:con}.

\section{Method and observed quantities}
\label{sec:model}
Monte Carlo simulations are carried out for the $J_x$-$J_y$ Blume-Capel model considered in Ref.~\cite{PH.A1}, using the standard Metropolis update. At each update, a lattice site is selected and its spin $S_i$ is proposed to change from its current state to one of the other allowed values $S_i=0,\pm1$. Each trial state is accepted with the probability
\begin{equation}
P_{\mathrm{acc}} = \min \left[ 1, \exp\left( -\frac{\Delta E}{k_{\mathrm{B}} T} \right) \right],
\label{eq:pacc}
\end{equation}
where $\Delta E$ is the energy difference produced by the trial spin change.
In this section, we briefly introduce several primary quantities that serve as order parameters in this paper.
\begin{itemize}
\item The staggered magnetization \( M_a \), as an order parameter to capture the features of the SAF phase, is defined as
\begin{equation}
M_a=\frac{1}{N}\sum_{(x,y)}(-1)^x S_{(x,y)},
\end{equation}
where $x$ and $y$ denote the horizontal and vertical lattice coordinates, respectively, and $N=L^2$ is the total number of sites of an $L\times L$ lattice. The factor $(-1)^x$ captures the staggered ordering along the $x$ direction of the SAF state.
\end{itemize}
\begin{itemize}
\item The squared staggered magnetization is
\begin{equation}
m_a^2=\langle M_a^2\rangle.
\end{equation}
\end{itemize}
\begin{itemize}
\item The Binder ratio is defined as 
\begin{equation}
Q=\frac{\langle M_a^2\rangle^2}{\langle M_a^4\rangle}.
\end{equation}
\end{itemize}
\begin{itemize}
\item The specific heat $C_V$ is expressed as
\begin{equation}
C_V=\frac{N}{k_{\mathrm B}T^2}\left(\langle e^2\rangle-\langle e\rangle^2\right),
\end{equation}
where $e=E/N$ is the energy per site, so $C_V$ is the specific heat per site.
\end{itemize}
\begin{itemize}
\item The entropy $S$ is expressed as
\begin{equation}
S(T)=S_0+\int_0^T\frac{C_V(T')}{T'}\,dT',
\end{equation}
where $S_0=S(0)$ is the zero-temperature entropy per site and $C_V$ is the specific heat per site.
\end{itemize}

\section{Results}
\label{sec:res}
{In this section, we present the main results obtained from the Monte Carlo (MC) method and compare these findings with the conclusions established in \cite{PH.A1} and \cite{PH.A2}.}

\subsection{Phase diagram without magnetic field}
\begin{figure}[!t]
\centering
\includegraphics[width=\linewidth]{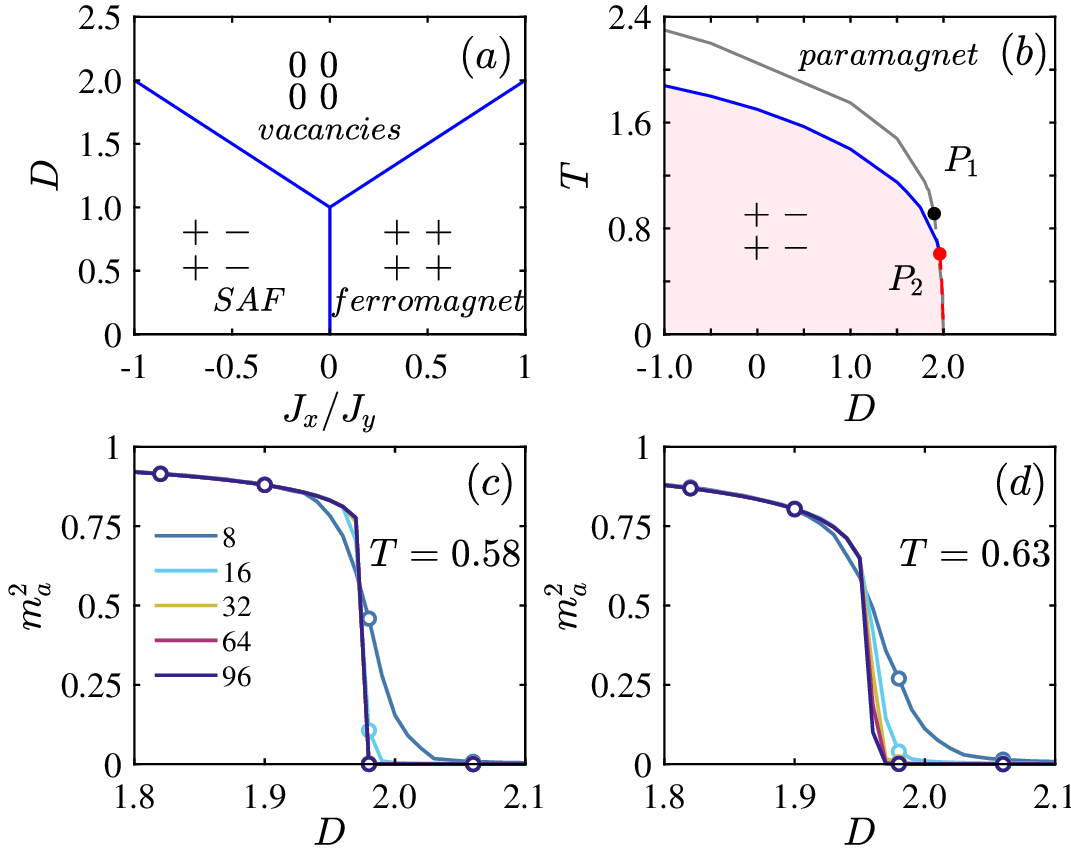}
\caption{Zero-field phase behavior of the $J_x$-$J_y$ Blume-Capel model with $J_y=1$. (a) Ground-state phases determined by $J_x/J_y$ and $D$. (b) $T$-$D$ phase diagram at $J_x=-1$ and $h=0$. The gray solid and dashed segments are the second- and first-order CV boundaries from Ref.~\cite{PH.A1}, while the blue solid and red dashed curves are the corresponding MC boundaries; the light shaded region denotes the MC SAF region, and $P_1$ and $P_2$ mark the CV and MC tricritical points, respectively. (c) and (d) $m_a^2$ as a function of $D$ at $T=0.58$ and $0.63$ for $L=8$, 16, 32, 64, and 96.}
\label{zero_field_phase}
\end{figure}

\begin{figure}[!t]
\centering
\includegraphics[width=\linewidth]{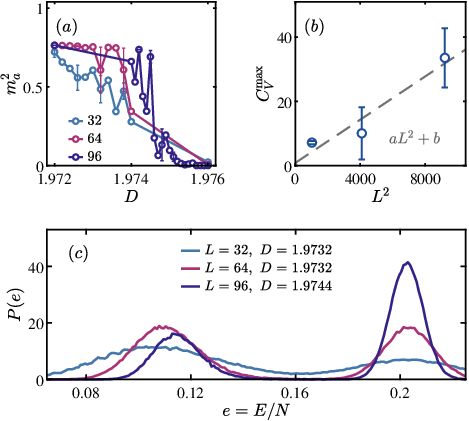}
\caption{Finite-size behavior near the zero-field transition at $T=0.58$. 
    (a) Crystal-field dependence of the squared antiferromagnetic order parameter $m_a^2$ for $L=32$, 64, and 96. 
    (b) Maximum specific heat $C_V^{\max}$ as a function of $L^2$. The gray dashed line, $aL^2+b$, is a guide to the $C_V^{\max}\propto L^2$ volume scaling expected for a first-order transition. 
    (c) Energy-density distributions $P(e)$, with $e=E/N$, evaluated at $D=1.9732$ for $L=32$ and 64 and at $D=1.9744$ for $L=96$.}
\label{zero_field_fss}
\end{figure}

In the absence of magnetic fields and at extremely low temperatures, this anisotropically interacting system exhibits three fundamental phases: ferromagnetic (F), vacancies (V), and a novel phase referred to as SAF~\cite{PH.A1}, which are illustrated in Fig.~\ref{zero_field_phase} (a). For small positive values of the parameter \(D\), and when \(J_x/J_y < 0\), the ground state is SAF; conversely, when \(J_x/J_y > 0\), the ground state can either be ferromagnetic (F) or antiferromagnetic (AF), contingent upon specific conditions, such as \(J_x = J_y = -1\) or \(J_x = J_y = 1\). For simplicity, we set \(J_y = 1\) in the following discussions. At high positive values of \(D\), the ground state corresponds to vacancies, where all sites are occupied by the 0-component of the spin. When \(D\) is negative, the ground state favors \(S_i = \pm 1\) and exhibits Ising-type behavior, although this is not elaborated upon here. The point \(J_x = 0\) represents a coexistence point of the F and SAF phases, while other boundaries can be determined analytically through the relationships outlined below:
\begin{align}
\begin{aligned}
& E_{\begin{smallmatrix} +-\\+- \end{smallmatrix}} = E_{\begin{smallmatrix} 00\\00 \end{smallmatrix}}, \\
& E_{\begin{smallmatrix} ++\\++ \end{smallmatrix}} = E_{\begin{smallmatrix} 00\\00 \end{smallmatrix}}, \\
& E_{\begin{smallmatrix} +-\\+- \end{smallmatrix}} = E_{\begin{smallmatrix} ++\\++ \end{smallmatrix}}.
\end{aligned}
\quad \Rightarrow \quad
\begin{aligned}
& J_x-J_y+D=0, \\
& -J_x-J_y+D=0, \\
& J_x=0.
\end{aligned}
\end{align}
In Fig.~\ref{zero_field_phase}(b), we examine the phase transitions from the SAF phase to other phases, which are influenced by both the crystal field parameter \(D\) and thermal effects \(T\). The gray line represents the results obtained from the CV method~\cite{PH.A1}, where the solid portion indicates the critical line for second-order transitions that separates the paramagnetic phase from the SAF phase. Conversely, the dashed portion delineates the first-order transition line that separates the vacancies from the SAF phase. These two lines converge at a tricritical point (TCP), indicated by the black dot. The blue solid line and red dashed line illustrate the findings from Monte Carlo (MC) simulations, corresponding to second-order and first-order phase-transition boundaries, respectively. Solid circles represent TCPs derived from different methodologies: \(P_1\), the CV result, is denoted by the black dot, while \(P_2\), the MC result, is denoted by the red dot. Notably, the critical temperature derived from the MC simulations is significantly lower than that predicted by the CV method, as evidenced by the position of the red dot in Fig.~\ref{zero_field_phase}(b).
Although we consider the anisotropic interaction with \( J_x = -J_y \), the location of the tricritical point remains unchanged. Figures \ref{zero_field_phase}(c) and (d) provide insights into the phase transitions near the tricritical point. We determine that the critical temperature lies within the range of \( T = 0.58 \) to \( T = 0.63 \). The abrupt drop of \( m_a^2 \) in Fig. \ref{zero_field_phase}(c) and the smooth decrease observed in Fig. \ref{zero_field_phase}(d), indicating the presence of two distinct categories of phase transitions around this region. Our findings are in good agreement with the results reported in~\cite{Wang-Landau}, where \( T_c = 0.609(3) \) and \( D_c = 1.966(2) \). This invariance arises from the fact that, for a square lattice, phase transitions are primarily influenced by symmetry rather than geometric factors and interactions. In addition to the location of the TCP, we notice that in the absence of an external magnetic field, mean-field approaches neglect thermal fluctuations. This results in a relatively larger predicted area of the ordered phase. In contrast, Monte Carlo simulations suggest that this area is in fact, smaller, as illustrated by the shaded region in Fig. \ref{zero_field_phase}(b).
The abrupt change in Fig.~\ref{zero_field_phase}(c) suggests that the transition at $T=0.58$ lies on the first-order side, and the finite-size behavior in Fig.~\ref{zero_field_fss} provides a more direct check. As $L$ increases, the change of $m_a^2$ becomes progressively sharper [Fig.~\ref{zero_field_fss}(a)], while the maximum specific heat increases approximately linearly with $L^2$ [Fig.~\ref{zero_field_fss}(b)]. The corresponding energy distributions also develop a clear double-peak structure [Fig.~\ref{zero_field_fss}(c)]. For $L=32$ and 64, the distributions are evaluated at $D=1.9732$, while that for $L=96$ is taken at $D=1.9744$. These finite-size features consistently support the first-order character of the zero-field transition at $T=0.58$.

\subsection{Magnetic field response}
\begin{figure}[!t]
\centering 
\includegraphics[width=\linewidth]{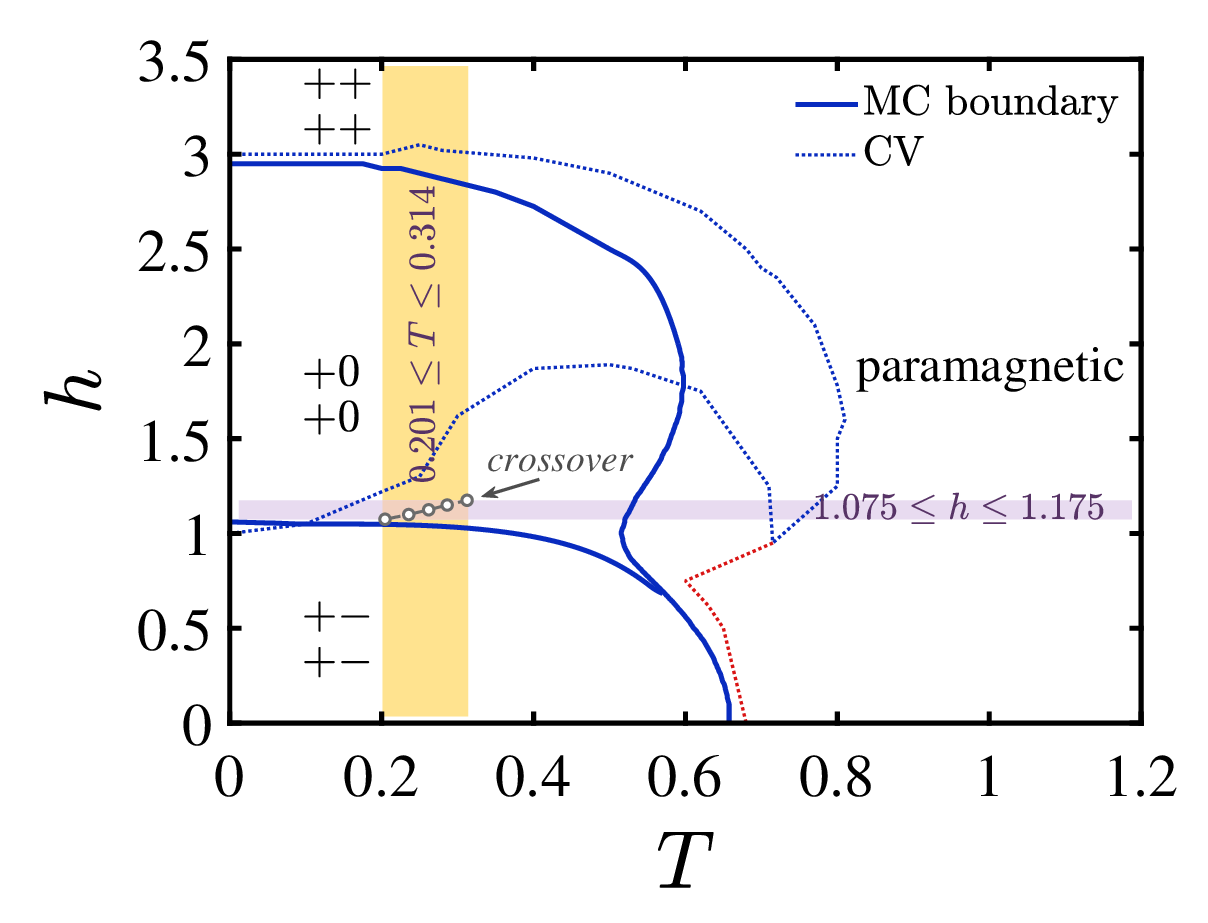}
\caption{Magnetic-field phase diagram at $J_x=-1$, $J_y=1$, and $D=1.95$. The blue solid curve denotes the MC phase boundary, while the blue and red dotted curves are the second- and first-order CV boundaries from Ref.~\cite{PH.A1}. Gray open circles trace the crossover points; the yellow and purple shaded bands indicate the temperature interval $0.201\leq T\leq0.314$ and field interval $1.075\leq h\leq1.175$ spanned by these points, respectively. The configurations $\begin{smallmatrix}+-\\+-\end{smallmatrix}$, $\begin{smallmatrix}+0\\+0\end{smallmatrix}$, and $\begin{smallmatrix}++\\++\end{smallmatrix}$ denote the SAF, mixed, and fully polarized states, respectively.}
\label{field_phase}
\end{figure}

In this section, we discuss the modifications of the SAF state $\begin{smallmatrix} +-\\+- \end{smallmatrix}$ when it is subjected to a external magnetic field. Generally, the magnetic field can regulate the spins, with a high external field $h$ favoring the F state or saturated paramagnetic state $\begin{smallmatrix} ++\\++ \end{smallmatrix}$. In contrast, a high crystal field $D$ promotes the vacancy state $\begin{smallmatrix} 00\\00 \end{smallmatrix}$, while elevated temperatures $T$ favors the paramagnetic state. For intermediate values of $D$, the system is in a mixed state $\begin{smallmatrix} +0\\+0 \end{smallmatrix}$, indicating that some lattice sites are occupied by the 0-component of the spin.
For $J_x=-1$, $J_y=1$, and $D=1.95$, which is also a parameter value utilized in~\cite{PH.A1}, the two zero-temperature reference fields can be obtained directly by comparing the energies per site of the SAF, mixed, and fully polarized configurations. Writing the energy per site and counting each nearest-neighbor bond once, the Hamiltonian in Eq.~\eqref{eq:ham1} gives
\begin{align*}
e_{\mathrm{SAF}} &= J_x-J_y+D=-0.05,\\
e_{\mathrm{mixed}} &= -\frac{J_y}{2}+\frac{D}{2}-\frac{h}{2}
=0.475-\frac{h}{2},\\
e_{\mathrm{pol}} &= -J_x-J_y+D-h=1.95-h.
\end{align*}
For the mixed configuration, the horizontal exchange contribution vanishes, while only half of the vertical bonds contribute and half of the sites carry $S=+1$, which gives the one-half factors in the vertical exchange, crystal-field, and Zeeman terms. The vacancy configuration has $e_{\mathrm{V}}=0$ and is slightly higher than the SAF state at $h=0$, so it does not enter the positive-field ground-state sequence for this parameter set. The first boundary is therefore obtained from $e_{\mathrm{SAF}}=e_{\mathrm{mixed}}$,
\begin{equation*}
-0.05=0.475-\frac{h_{c1}}{2},
\qquad h_{c1}=1.05,
\end{equation*}
while the second follows from $e_{\mathrm{mixed}}=e_{\mathrm{pol}}$,
\begin{equation*}
0.475-\frac{h_{c2}}{2}=1.95-h_{c2},
\qquad h_{c2}=2.95.
\end{equation*}
The direct SAF--polarized energy crossing occurs at $h=2$, but the mixed state has lower energy there and therefore remains the ground state. Thus, at $T=0$ the system changes from SAF to mixed at $h_{c1}$ and from mixed to the fully polarized state at $h_{c2}$. 
Further details at finite temperature can be found in Fig.~\ref{field_phase}.

\subsubsection{The first-order phase transition}
Firstly, we examine the transition around $T=0.63$ at $h=0$ and $D=1.95$. In the CV result of Ref.~\cite{PH.A1}, this point lies on the first-order boundary (the red dotted line in Fig.~\ref{field_phase}), where the SAF state $\begin{smallmatrix} +- \\ +- \end{smallmatrix}$ is predicted to transform directly into the vacancy state $\begin{smallmatrix} 00 \\ 00 \end{smallmatrix}$ as the temperature increases. Our MC result gives a different picture: the corresponding boundary remains on the continuous branch and the SAF order evolves into the paramagnetic phase.
\begin{figure}[!t]
\centering
\includegraphics[width=\linewidth]{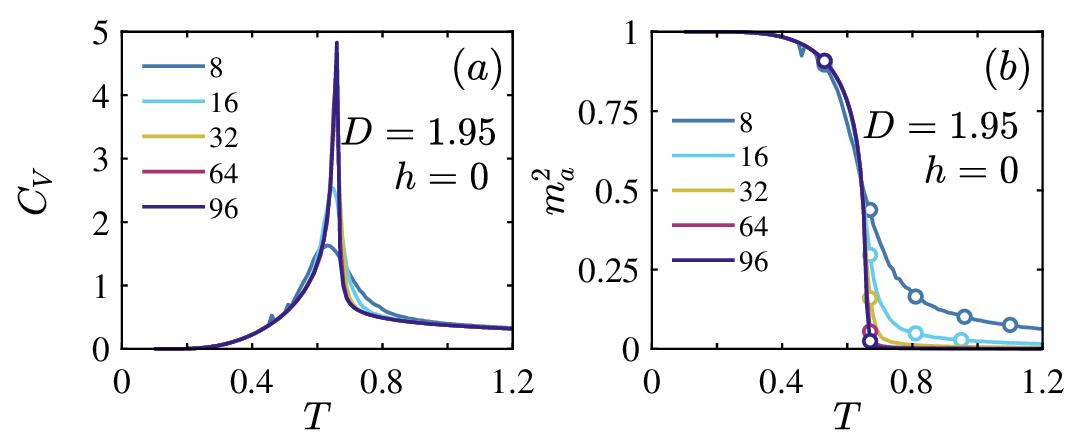}
\caption{Finite-size thermal behavior at $J_x=-1$, $J_y=1$, $D=1.95$, and $h=0$. (a) Specific heat $C_V$ as a function of temperature for $L=8$, 16, 32, 64, and 96. (b) Squared antiferromagnetic order parameter $m_a^2$ for the same system sizes.}
\label{zero_field_transition}
\end{figure}
The finite-size behavior is shown more directly in Fig.~\ref{zero_field_transition}. As the system size increases, the $C_V$ peak becomes sharper and remains centered near $T\simeq0.66$ [Fig.~\ref{zero_field_transition}(a)], while $m_a^2$ decreases continuously through the same temperature region [Fig.~\ref{zero_field_transition}(b)]. These two quantities therefore give a consistent transition temperature and support the continuous SAF-to-paramagnetic transition obtained from the MC phase boundary.

\subsubsection{The mixed phase and reentrance}
\begin{figure}[!t]
\centering
\includegraphics[width=\linewidth]{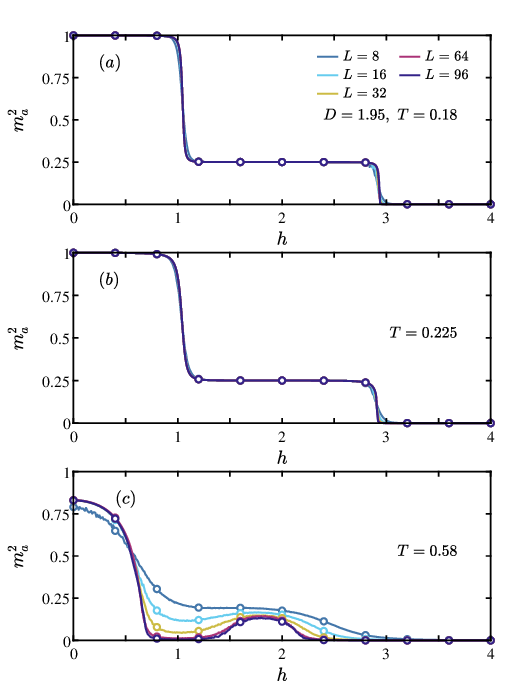}
\caption{Magnetic-field dependence of $m_a^2$ at $J_x=-1$, $J_y=1$, and $D=1.95$. Panels (a), (b), and (c) correspond to $T=0.18$, $0.225$, and $0.58$, respectively, with $L=8$, 16, 32, 64, and 96. At the two lower temperatures, the plateau near $m_a^2=1/4$ characterizes the mixed state, while the plateau becomes less distinct at $T=0.58$.}
\label{field_order}
\end{figure}

The low-temperature magnetic-field dependence of $m_a^2$ is shown in Fig.~\ref{field_order}(a) and (b), for temperatures close to $T=0.2$ at $D=1.95$. Notably, a plateau in $m_a^2 = 1/4$ is observed, which serves as an indicator of the mixed state $\begin{smallmatrix} +0 \\ +0 \end{smallmatrix}$.
Reentrant behavior refers to the process by which a system transitions from a specific ordered state to another one, and then reverts back to the original, as one parameter varies.
In Fig.~\ref{field_phase}, the CV result shows an interesting phenomenon where the system experiences phase transitions first from a mixed phase$\begin{smallmatrix} +0 \\ +0 \end{smallmatrix}$ to the SAF phase $\begin{smallmatrix} +- \\ +- \end{smallmatrix}$, and then back to the mixed phase as the temperature continues to increase. However, our MC calculation confirmed that temperature-dependent reentrant behavior could not be identified under the same parameter regime. Reentrant behavior only occurs as the magnetic field varies, and the temperature should be held constantly, as summarized in Fig.~\ref{field_order}(c).

\subsubsection{The spontaneous symmetrybreaking phase}
\begin{figure}[!t]
\includegraphics[width=\linewidth]{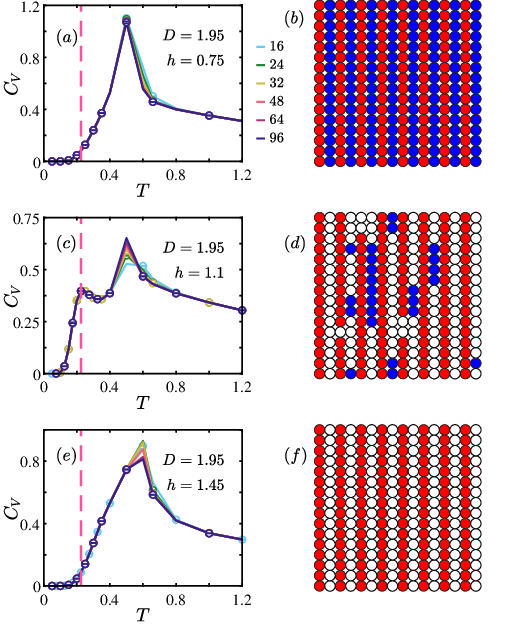}
\caption{Thermal response at $J_x=-1$, $J_y=1$, and $D=1.95$. Panels (a), (c), and (e) show $C_V(T)$ at $h=0.75$, $1.1$, and $1.45$, respectively, for $L=16$, 24, 32, 48, 64, and 96; the magenta dashed line marks $T=0.225$. Panels (b), (d), and (f) are the corresponding $L=16$ spin configurations at $T=0.225$ for $h=0.75$, $1.1$, and $1.45$, respectively. Red, blue, and open circles denote $S=+1$, $-1$, and $0$, respectively.}
\label{thermal_response}
\end{figure}

In the mixed state region (specifically at $h=1.1$), we found an unexpected phenomenon: a non-divergent turning point in the specific heat. We observed an extraordinary phase transition signature, characterized by a hump of $C_V$ within a certain temperature range, Fig.~\ref{thermal_response}(c). The typical SAF state 
is shown in Fig.~\ref{thermal_response}(b) , which can be interpreted as two antiferromagnetically interacting necklace.
The mixed state is Fig.~\ref{thermal_response}(f), which
can be attributed to the formation of 0-component layers, situated between two layers of non-zero spins. Consequently, this arrangement transforms the square lattice configuration into many one-dimensional chains. Within each chain, the comparable competition between $h$ and $D$ leads to the formation of the intermediate state, see Fig.~\ref{thermal_response}(d). This  state has another interpretation in the filed of spin-1 condensate system,  called spontaneous symmetry‑broken state ~\cite{Stenger1998,Sadler2006,Vinit2013}. Here, the term ``spontaneous'' means that sudden quenching of the magnetic field can lead to the emergence of total spin Larmor precession in a lower magnetic field intensity.

\begin{figure}[!t]
\centering
\includegraphics[width=0.65\linewidth]{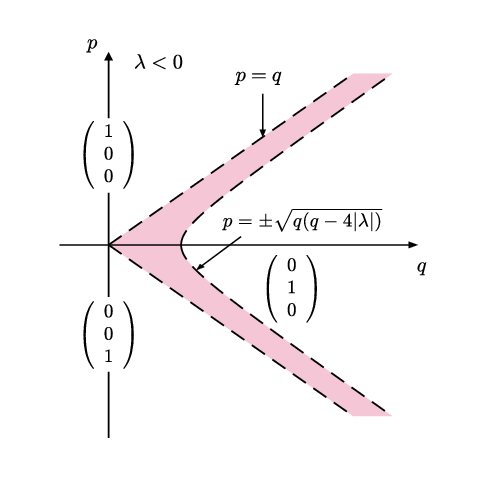}
\caption{Spin-domain diagram for the ferromagnetic $F=1$ spinor-condensate case, with the interaction parameter $\lambda<0$ ~\cite{Stenger1998}. The horizontal and vertical axes denote the quadratic and linear Zeeman parameters $q$ and $p$, respectively. The shaded regions contain mixed hyperfine components. The dashed boundaries indicate gradual changes in the state populations, with the corresponding boundaries given by $p=\pm q$ and $p=\pm\sqrt{q(q-4|\lambda|)}$.}
\label{spin_domain}
\end{figure}

The relation between $h$ and $D$ can be further understood by comparing the present model with the spin-domain picture of Rb condensate in Ref.~\cite{Stenger1998}. For an $F=1$ spinor condensate, the spin-dependent free energy under mean‑field approximation can be written as
\begin{equation}
K_s=\lambda\langle\mathbf{F}\rangle^2-p\langle F_z\rangle+q\langle F_z^2\rangle ,
\end{equation}
where $\lambda$ denotes the spin-dependent interaction~\cite{Stenger1998}, while $p$ and $q$ denote the linear and quadratic Zeeman contributions, respectively. The last two terms have the same single-component structure as the crystal-field and magnetic-field terms here
\begin{equation}
p\leftrightarrow h,\qquad q\leftrightarrow D.
\end{equation}
In Fig.\ref{spin_domain}, we recall the spin-domain construction in Ref.~\cite{Stenger1998} for the ferromagnetic $\lambda<0$ case. 
In the present BC model, the corresponding $D$--$h$ dependence can be understood analogously.  We give a crossover region in Fig.~\ref{field_phase}, to show possible regime to capture this special state.

It is worth emphasizing that, constrained by the transverse antiferromagnetic interaction $J_x$, our crossover regime is not fully equivalent to the above‑mentioned Larmor precession state. The magnetic field $h$ ramped up from zero tends to destabilize the vacancy and generate $S$=+1 or $S$=-1. However, the constraint in the transverse ($J_x$ interaction) suppresses the $S=+1$ component, which finely produce the only $S$=-1 component in the vacancy-lines, see the blue dots in Fig.~\ref{thermal_response}(d). The small hump of $C_V$ in Fig.~\ref{thermal_response}(c) can also be understood from this configurational competition. In this crossover region, part of the thermal energy is involved in rearranging the $S$=0 and $S$=-1 components and their exchange connections, which enhances the energy fluctuations and  the specific heat. The broad and non-divergent character of this hump is therefore more naturally associated with enhanced configurational fluctuations than with an additional phase transition.

Here we can see a significant advantage of the MC method, the ability to provide direct visualizations of the system's configuration, and facilitate the understanding of its fundamental features. We can interpret the mixed state not only through the mathematical description $m_a^2=1/4$, but also by observing the distribution of vacancy sites.

\FloatBarrier
\subsubsection{The Magnetocaloric Effect}
We next examine the magnetocaloric effect (MCE) associated with the field-induced changes discussed above. For a magnetic-field increase from $0$ to $h$, the isothermal magnetic entropy change is evaluated from the Maxwell relation
\begin{equation}
\Delta S_m(T,h)=\int_0^h
\left(\frac{\partial m}{\partial T}\right)_{h'}dh',
\end{equation}

\begin{figure}[!t]
\includegraphics[width=\linewidth]{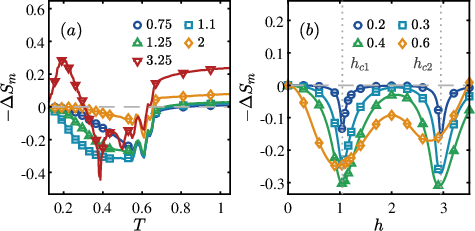}
\caption{Magnetocaloric response for $D=1.95$ and $J_x/J_y=-1$. (a) Isothermal magnetic entropy change $-\Delta S_m$ as a function of temperature for $h=0.75$, $1.1$, $1.25$, $2$, and $3.25$. (b) $-\Delta S_m$ as a function of magnetic field for $T=0.2$, $0.3$, $0.4$, and $0.6$. The horizontal dashed line marks $-\Delta S_m=0$. The vertical dotted lines in panel (b) mark the zero-temperature reference transition fields $h_{c1}=1.05$ and $h_{c2}=2.95$ discussed above.}
\label{mce_D195}
\end{figure}

Where $m$ is the uniform magnetization per site. We present the results as $-\Delta S_m$, so that $-\Delta S_m>0$ corresponds to the direct MCE and $-\Delta S_m<0$ to the inverse MCE. Since $\Delta S_m(T,h)$ contains the response accumulated over the field interval from $0$ to $h$, its sign should be understood as an integrated thermodynamic response rather than as an order parameter of the state at the final field.
We first consider $D=1.95$ and $J_x/J_y=-1$, so that the magnetocaloric response can be compared directly with the field-dependent phase behavior discussed in the preceding subsection. 
Figure~\ref{mce_D195}(a) shows that both the sign and magnitude of the MCE change considerably with magnetic field. Around the onset of the mixed-state field range, especially for $h=1.1$ and $1.25$, a pronounced inverse response develops at low and intermediate temperatures. At higher fields the temperature dependence changes again, indicating that the entropy response is sensitive to the field-driven rearrangement of the competing spin configurations.

The field scans in Fig.~\ref{mce_D195}(b) show the same behavior more directly. At low temperatures, negative structures develop around the two reference fields $h_{c1}=1.05$ and $h_{c2}=2.95$, which correspond to the ground-state boundaries separating the SAF, mixed, and polarized sectors discussed above. As the temperature increases, these features become broader and the distinction between the corresponding field ranges gradually weakens. The strongest inverse response therefore appears in the same low-temperature field region where the mixed-state behavior becomes important, although the integrated character of $\Delta S_m$ prevents its sign from being assigned uniquely to a single phase.

\begin{figure}[!t]
\includegraphics[width=\linewidth]{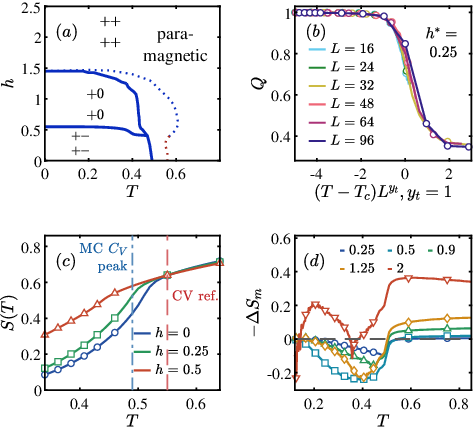}
\caption{Magnetocaloric response and transition diagnostics for $D=1.45$ and $J_x/J_y=-0.5$. (a) $T$-$h$ phase boundaries from MC and the CV calculation of Ref.~\cite{PH.A2}. The blue solid curve denotes the MC boundary, while the blue dotted curve and the short red dotted segment denote the second- and first-order CV boundaries, respectively. (b) Binder-ratio data collapse at $h^*=0.25$ for $L=16$, 24, 32, 48, 64, and 96 using $y_t=1$. (c) Entropy $S(T)$ for $h=0$, $0.25$, and $0.5$; the blue and red vertical lines mark the MC specific-heat peak and the CV first-order reference temperature, respectively. (d) $-\Delta S_m(T)$ for $h=0.25$, $0.5$, $0.9$, $1.25$, and $2$.}
\label{mce_D145}
\end{figure}

Finally, we consider a second parameter set to compare more directly with the magnetocaloric results in Ref.~\cite{PH.A2}. For $J_x/J_y=-0.5$ and $D=1.45$, the CV calculation predicts a low-field first-order sector together with a discontinuous entropy response. This makes the same parameter region useful for checking whether these features remain after the phase boundary is recalculated by MC.
Figure~\ref{mce_D145}(a) shows a clear difference between the MC and CV phase boundaries. The MC boundary is shifted toward lower temperatures and does not reproduce the short first-order segment obtained from the CV calculation. At the representative field $h^*=0.25$, the Binder-ratio data for $L=16$--$96$ collapse reasonably well with $y_t=1$ [Fig.~\ref{mce_D145}(b)], supporting a continuous thermal transition at this point within the present finite-size analysis.

The entropy curves in Fig.~\ref{mce_D145}(c) give a consistent picture. They remain smooth through the transition range, and no finite entropy jump is resolved around either the MC specific-heat peak or the first-order reference temperature obtained from the CV result. Correspondingly, the magnetic entropy changes in Fig.~\ref{mce_D145}(d) also vary continuously with temperature. Both direct and inverse MCEs remain present and their relative amplitudes change strongly with field, but the discontinuous entropy response associated with the CV first-order boundary is not observed within the present numerical resolution.

The two parameter sets therefore probe complementary aspects of the magnetocaloric behavior. For $D=1.95$ and $J_x/J_y=-1$, the sign-changing MCE follows the same field scale over which the SAF, mixed, and polarized configurations compete. For $D=1.45$ and $J_x/J_y=-0.5$, the calculation directly tests the first-order magnetocaloric scenario proposed in Ref.~\cite{PH.A2}. The MC results retain both direct and inverse responses, while the phase boundary, Binder-ratio scaling, and smooth entropy behavior do not support the discontinuous response expected from the CV first-order sector.

\FloatBarrier
\section{Conclusions}
\label{sec:con}
In this paper, we use Monte Carlo (MC) simulations to reexamine the finite-temperature phase behavior and magnetocaloric response of the $J_x$-$J_y$ Blume-Capel model, with particular attention to the cluster-variation (CV) results reported in Refs.~\cite{PH.A1,PH.A2}. The MC results retain the basic SAF and field-induced mixed-state structure, but important differences appear in the phase boundaries and in the character of several transitions.

\noindent
(i) At zero magnetic field, the ordered region obtained from MC is clearly smaller than that given by the CV calculation of Ref.~\cite{PH.A1}, while the location of the tricritical region remains close to the conventional square-lattice BC result. This difference should not be attributed simply to an absence of thermal fluctuations in the CV method. Rather, the finite-cluster approximation treats correlations differently from the lattice MC simulation and, in the present case, gives a higher stability of the ordered phase.

\noindent
(ii) For $h=0$ and $D=1.95$, Ref.~\cite{PH.A1} places the corresponding thermal transition on the first-order side of the CV phase boundary. Our MC result instead gives a continuous SAF-to-paramagnetic transition, with the finite-size $C_V$ and $m_a^2$ results in Fig.~\ref{zero_field_transition} locating the transition near $T\simeq0.66$. At the same time, the finite-size analysis at $T=0.58$ in Fig.~\ref{zero_field_fss} shows the sharpening order-parameter change, the growth of the specific-heat maximum with system size, and a double-peaked energy distribution expected for a first-order transition. Thus, the discrepancy mainly concerns the location of the first- and second-order sectors rather than the disappearance of first-order behavior from the model.

\noindent
(iii) Under a magnetic field, the intermediate mixed state remains a robust feature of the MC results and is identified by the plateau near $m_a^2=1/4$ together with the corresponding vacancy-rich configurations. The temperature-driven reentrant sequence predicted by the CV phase diagram is not reproduced in the same parameter range. The MC phase boundary nevertheless remains strongly nonmonotonic, showing that the competition among the SAF, mixed, and polarized configurations is still important.

\noindent
(iv) For $D=1.95$ and $J_x/J_y=-1$, both direct and inverse magnetocaloric responses are obtained. The zero-temperature energy comparison gives $h_{c1}=1.05$ and $h_{c2}=2.95$ for the SAF--mixed and mixed--polarized boundaries, respectively. The strongest inverse response develops in the same low-temperature field range around these field-driven rearrangements. Since $\Delta S_m$ is integrated from zero field to the final field, its sign cannot by itself be assigned to a single equilibrium phase, but the MCE still provides a thermodynamic signature of the same field-induced competition.

\noindent
(v) Ref.~\cite{PH.A2} predicts, for $J_x/J_y=-0.5$ and $D=1.45$, a tricritical and reentrant structure together with discontinuous entropy changes associated with the first-order sector. Our MC results do not reproduce the short first-order boundary in this region. The Binder-ratio scaling at $h^*=0.25$ is consistent with a continuous transition, while the entropy and magnetic entropy-change curves remain smooth through the corresponding temperature range. Direct and inverse MCEs are still present, but no discontinuous entropy response associated with the CV first-order boundary is resolved within the present numerical accuracy.

The most important finding is that, between SAF- and mixed-states, there is a crossover region where the competing vacancy-rich and non-zero-spin configurations can develop a spontaneous spatial organization. The accompanying configurational rearrangement enhances the energy fluctuations and gives rise to the broad, non-divergent specific-heat hump, which can be interpreted as a crossover feature rather than evidence for an additional phase transition. Compared with conventional solid materials, the BC model now can be explored in systems that provide highly controllable magnetic fields and low temperatures, even the crystal-field term $D$ can be tuned through microwave manipulation~\cite{LiYou}. Together with optical-lattice techniques or artificial
spin ice experiments, our work provides a pioneering exploration for further studies.

\section*{Acknowledgments}
\begingroup
\tolerance=5000
\emergencystretch=2em

\par
\endgroup
\bibliography{ref}
\end{document}